\documentclass[MSNbibl,nameyear,dvips]{arxstspdf}
\usepackage{graphicx}
\usepackage{flushend}
\usepackage{stfloats}

\volume{26}
\issue{1}
\pubyear{2011}
\firstpage{59}
\lastpage{61}
\doi{10.1214/11-STS345REJ}
\referstodoi{10.1214/10-STS345}

\begin{document}
\begin{frontmatter}

\vspace*{12pt}
\title{Rejoinder}
\runtitle{Rejoinder}
\pdftitle{Rejoinder of Feature Matching in Time Series Modeling by Y. Xia and H. Tong}

\begin{aug}
\author{\fnms{Yingcun} \snm{Xia}\corref{}\ead[label=e1]{staxyc@nus.edu.sg}}
\and
\author{\fnms{Howell} \snm{Tong}\ead[label=e2]{howell.tong@gmail.com}}
\runauthor{Y. Xia and H. Tong}

\affiliation{National University of Singapore and London School of Economics}

\address{Yingcun Xia is Professor of Statistics, Department of Statistics and Applied
Probability, and Risk Management Institute,
National University of Singapore, Science Drive~2, Kent Ridge,
Singapore 117546 \printead{e1}.
Howell Tong is Emeritus Chair Professor of Statistics, London School of Economics,
Houghton St, London WC2A 2AE,
United Kingdom \printead{e2}.}

\end{aug}



\end{frontmatter}

We would like to thank all the  discussants for their wide-ranging
comments. Before we respond to them individually in alphabetical
order, we would like to address some general issues first. As we
have said, we have chosen to describe our aim of matching the joint
distribution of the observable data as \textit{feature matching}, for
want of a better name. We should have perhaps emphasized that we
regard cycles, spectral singularities, and so on only as partial aspects of
the joint distribution. They are useful, in practical applications,
only in so far as they can provide partial measures of feature
matching. We think Professor Hansen has understood our aim well in
his introduction. We have sometimes, for brevity, called our general approach to achieving
this aim the \textit{catch-all approach}. We should stress the following
point once more. The catch-all approach is not restricted to catch all
first-order (conditional) moments or catch all second-order moments.
We have used them in the paper primarily as illustrations of what the
approach can deliver in modeling, beyond conventional methods based
on the one-step-ahead prediction errors. Clearly, once the catch-all idea
is accepted,  we can equally well catch all
$k$th-order (conditional or unconditional) moments, catch all marginal
(conditional or unconditional) distributions, and so on. Moreover,
the objective function $Q$ can also take on a~form other than that of a
mean squared type; for example, it can be of a likelihood type as
stated in Section~2.1.

Professors Chan and Tsay have tried the catch-all approach on two
real data sets, namely (i) the CREF stock fund and (ii) the monthly
global temperature anomalies from 1880 to 2010. In each case, their
implementation of the approach is exemplary. In data set (i), the
catch-all approach has led to parameter estimates of the postulated
$\operatorname{GARCH}(1,1)$ model that enable the model to ``track the squared
returns more closely'' and ``transit into the ensuing quiet period at
a faster rate commensurate with the data.'' We are sure that Chan and
Tsay are aware of the fact that the larger is $\alpha$, the more
responsive is the $\operatorname{GARCH}(1,1)$ model to volatility.

Chan and Tsay seem to be disappointed with their attempt with data
set (ii). They have correctly noted the shapes of the eventual
forecasting functions (eff) of the $\operatorname{ARIMA}(1,1,1)$ model and the
$\operatorname{ARMA}(1,1)$-plus-trend model. Now, long-range forecasting invokes a~low pass filter, which is approximately provided by the eff.
Therefore, for an $\operatorname{ARIMA}(1,1,1)$ model, for sufficiently large $ l$
and conditional on~$Y_s$,  $s \le t$, $EY_{t + l} \approx
K$, where $K$ is a constant. In such cases, $\phi \approx -\theta$,
the well-known near cancelation of the AR operator and the MA
operator. Similar arguments apply to an $\operatorname{ARMA}(1,1)$ model. It is clear
in the setup of Chan and Tsay, as $m$ increases, long-range
forecasts exert greater and ultimately overwhelming influence on the
objective function, $S$.\break Thus, evidence of operator near cancelation
with increasing $m$ is evidence of plausibility of the postulated
model. This argument suggests that if Chan and Tsay had perhaps
probed further with their Figure 2, they might be marginally more
inclined toward the $\operatorname{ARMA}(1,1)$-plus-trend model. Of course, we must
always be very cautious if we entertain any thought of extrapolating
the trend into the future.

\begin{figure}

\includegraphics{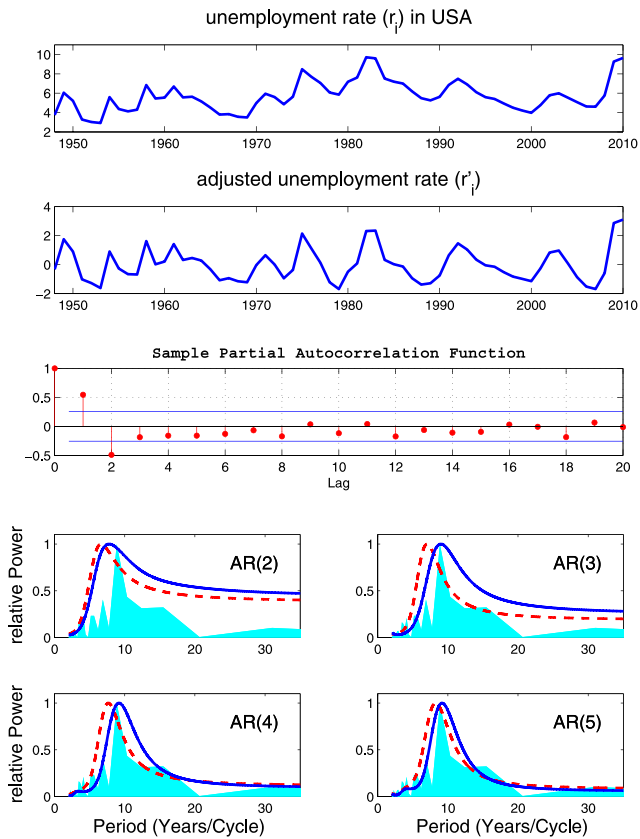}

\caption{The first panel is the unemployment rate
($r_i$) in the United States. The second panel adjusts it to  $ r_i'
= r_i - (r_{i-5}+r_{i-4} + \cdots + r_{i+5})/11$. Panel 3 is the plot
of partial autocorrelation function with approximate 95\% confidence
interval.  In each of panels 4--7, the edge of the shaded area is the
power spectrum of time series $ \{ r_i'\} $. The red dashed line and
blue solid line are those of the models estimated
 by $\operatorname{APE}(1)$ and
$\operatorname{APE}({<}T)$, respectively. }
\label{uerfig}
\end{figure}

Taking up the challenge posed by Chan and Tsay relating to business
cycles, we have considered the unemployment rate in the United States. The
second panel of Figure \ref{uerfig} shows the rate after the removal
of a moving mean. The partial autocorrelation function suggests
strong $\operatorname{AR}(2)$ effect with a hint of higher order dependence. Figure
\ref{uerfig} compares the spectral density functions of the $\operatorname{AR}$
model, from order 2 to 5, fitted respectively by the catch-all
approach and the maximum likelihood approach. The former approach
seems to show an overall better matching of the observed. The
fundamental period of 9 years is clearly discernible and reasonably
well captured by the $\operatorname{AR}(3)$, $\operatorname{AR}(4)$
and $\operatorname{AR}(5)$  models fitted by the
catch-all approach. The possible existence of higher harmonics
deserves further investigation, however.

Professor Hansen has made numerous perceptive comments. We are much
heartened by his endorsement of our most serious criticisms
of what we now call the \textit{unreasonableness} of one-step-ahead-predic\-tion-error based
methods. By paraphrasing Chan and Tsay, these methods have
been used as if they were ``one-size-fits-all'' for far too long.
Like the ancient practice of \textit{foot-binding} among Chinese women, they are constrictive and painful. We
advocate \textit{foot-unbinding}. Needless to say, we are not so naive
as to overlook the initial pains in unbinding or so arrogant as to
rule out the possibility of other ways to unbind.

We are also very grateful to Professor Hansen for furthering
the spirit of our Theorem C. We are in broad agreement with his
analysis. Our only quibble is with his reference (also raised by
Professor Ionides) to efficiency. For a wrong model, the
conventional notion of efficiency can be misleading. White (\citeyear{Whi82})
is relevant. We agree with Professor Hansen that there are many
troubling problems with measurement errors. Our own contributions
are quite modest in comparison with the enormity of the problems.
Even rounding the data can be very troublesome already. See, for example,
Zhang, Liu and Bai (\citeyear{ZhaLiuBai09}).

Professor Ionides is clearly a faithful adherent to the
maximum likelihood doctrine. Box's dictum tells us that all models are wrong.
Although some of them might be useful, they are still wrong. (Our paper does
not address model selection. We shall return to this point later.)
For a wrong model, what do we mean by  efficiency or
consistency? How would we assess likelihood ratios or AIC? Conventional
treatments are characteristically invalid in their original forms. For example,
we have highlighted the loss of a minimal set of sufficient
statistics in Section 1.2. Professor Ionides must accept that
the loss has by and large rendered the maximum likelihood type of estimation
impotent. Next, for a wrong model, the Hessian matrix
will typically have an expectation not equal to the negative of the
variance of the score matrix. This clearly has implications
on efficiency. As a third example, in the absence of a true
parameter, the notion of consistency will have to
be re-defined. It is precisely for this purpose that we have proposed the notion of a
($w$-dependent) optimal parameter. (By taking the infimum with
respect to $w$, we can also define the notion of an optimal
parameter.) Other examples abound. We would suggest that
it is high time that we unbound our feet.

\begin{figure}[t!]

\includegraphics{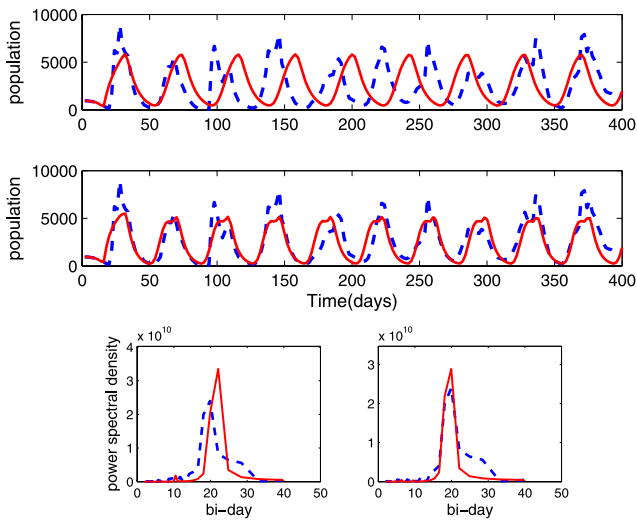}

\caption{Results of fitting the new model to Nicholson's blowflies data. In the
first two panels, the dashed lines are for the observed population; the
solid lines are for realizations from  models fitted by $\operatorname{APE}({\le}1$)
and $\operatorname{APE}({\le}T$), respectively. The dashed lines in panels~3 and 4 are the
periodograms of the observed data, and the solid
lines are those of the models fitted by $\operatorname{APE}({\le}1$) and $\operatorname{APE}({\le}T$),
respectively.
}\label{blowfig}
\end{figure}

The critical re-examination by Professor Ionides of the substantive model
that we have fitted to the blowfly data has revealed an interesting situation.
While the catch-all approach produces parameter estimates that lead
to a good statistical fit to the observed data, they are not
scientifically plausible. On the other hand, while the
maximum likelihood type estimates of the parameters of the same
model lead to a poor statistical fit, they are scientifically
plausible. We take the view that this apparent dichotomy suggests
that even a substantive model is not sacrosanct; an ideal model should
be, at least, satisfactory both statistically and scientifically.
Indeed, prompted by the alternative substantive model suggested by Ionides, we
have considered the following simple equivalent of his model by
merging his unobservable $R_t$ and $S_t$ into the available bi-daily
data:
\[
x_t = \operatorname{Poisson}\bigl(cx_{t-\tau}\exp(-x_{t-\tau}/N_0) + \nu x_{t-1}\bigr).
\]
The maximum likelihood estimates (with Poisson distribution) are
\[
c = 8.49,\quad  N_0=528.23,\quad  \nu = 0.77
\]
and the catch-all estimates (with the objective function based on
all-step-ahead predictors) are
\[
c = 8.82,\quad  N_0 = 604.98,\quad  \nu = 0.67.
\]
Both sets of parameter estimates are broadly consistent with those
obtained by Professor Ionides for his new model ($c = 2 \times 3.28,
N_0 = 680,  \nu = \exp(-2\delta) = 0.7247$), and lead to scientifically
plausible models ($c$~being the reproducing rate). On the other
hand, Figure~\ref{blowfig} shows that the catch-all approach gives a~%
good fit to the observed periodicity
but the maximum likelihood approach does not. Similar remark applies
to the fitted period as a function of time to maturity (not shown
as the results are similar to panels 5 and 6 of Figure 8 in the paper).
It is unfortunate that
results of the statistical goodness of fit
of his alternative model are not available.

Professor Yao is clearly alert and has noted our fleeting reference
to the important issue of model selection among wrong models. He has
given some valuable thoughts on the issue, for which we are most
grateful. We suspect that he will also agree with us that the
general problem is much deeper. The reference to Konishi and
Kitagawa (\citeyear{KonKit96}) is apt. As for his quibble, if he can throw us a
more catchy line than catch-all, we would be happy to catch it.
Perhaps, ``foot-unbinding'' might be a better name in reflecting the
philosophy of our approach.


\end{document}